\documentclass[aps,prb,reprint,showpacs,twocolumn,letterpaper,superscriptaddress,floatfix]{revtex4-1}

\usepackage{graphicx,subfigure}
\usepackage{amsmath,amssymb}
\usepackage{natbib}
\usepackage{bm}
\usepackage{color}
\newcommand{\be}{\begin{equation}}
\newcommand{\ee}{\end{equation}}
\newcommand{\ba}{\begin{eqnarray}}
\newcommand{\ea}{\end{eqnarray}}

\begin{document}
\title{Lefschetz thimble Monte Carlo for many body theories: application to the repulsive Hubbard model away from half filling}
\author{Abhishek Mukherjee}
\email[]{mukherjee@ectstar.eu, abhi.mukh@gmail.com}
\affiliation{ECT$^\star$, Villa Tambosi, I-38123 Villazzano (Trento), Italy}
\affiliation{LISC, Via Sommarive 18, I-38123 Povo (Trento), Italy}
\author{Marco Cristoforetti}
\email[]{mcristofo@ectstar.eu}
\affiliation{ECT$^\star$, Villa Tambosi, I-38123 Villazzano (Trento), Italy}
\affiliation{LISC, Via Sommarive 18, I-38123 Povo (Trento), Italy}

\begin{abstract}
Recently, a new method, based on stochastic integration on the surfaces of steepest descent of the action, was introduced to
tackle the sign problem in quantum field theories. We show how this method can be used in many body theories 
to perform fully non-perturbative calculations of quantum corrections about mean field solutions. We discuss an explicit algorithm 
for implementing our method, and present results for the repulsive Hubbard model away from half-filling at intermediate temperatures.
Our results are consistent with those from the recent state of the art cluster dynamical mean field theory calculations.
\end{abstract}
\maketitle

\section{Introduction}
In numerous physical systems, quantities of interest can be written as integrals over high dimensional distributions. 
When these distributions are positive semi-definite, then the most efficient and accurate method to evaluate them
is via stochastic (Monte Carlo) sampling. However, this is often not the case, leading to the infamous
\emph{sign problem}. The sign problem severely limits the usefulness and applicability of Monte Carlo
methods in many important systems. It is, without a doubt, one of the most important outstanding problems in computational science.

In previous publications we have shown how the same ideas used in constructing asymptotic approximations of oscillatory 
integrals can prove useful in  non-perturbative Monte Carlo calculations in the presence of a sign
problem \cite{Cristoforetti2012,Cristoforetti2013,Mukherjee2013,Cristoforetti2013a,Cristoforetti2013b}. 
The basic proposal is to change the domain of integration to the hypersurface generated by the paths of 
steepest descent of the action ending at one of the saddle points. On this hypersurface, the so called Lefschetz thimble,
the sign problem is either absent or significantly softer. We showed that for quantum field theories with a unique continuum limit
this procedure should yield results which are identical to those in the original formulation.

In this paper we discuss how the method can be extended for fermionic many body theories,
which do not necessarily have a unique continuum limit. We show that, within the functional integral formulation,
an integration on the said Lefschetz thimble(s) is a way to include all quantum corrections about one or more 
mean field solution(s) in a \emph{fully non-perturbative} manner.

We then apply this method to the two dimensional Hubbard model \cite{Hubbard1963,*Gutzwiller1963,*Kanamori1963}.
 The Hubbard model is the quintessential model of electronic correlations \cite{Nature2013}.  
It has been hypothesized to contain the essential physics of high temperature superconductivity (high $T_c$) \cite{Anderson1987}. 
More recently the Mott insulator phase of this model was realized experimentally in optical lattices with cold atoms \cite{Jordens2008}
and efforts are underway to realize other parameter regimes as well with these analog simulators \cite{Esslinger2010,*Bloch2012}. 
The small doping regime of this model, believed to 
be relevant for high $T_c$, is least well understood theoretically because of the presence of the sign problem \cite{Loh1990,Loh1992,Bai2009}. 

In the next section we briefly discuss the functional integral formulation via Hubbard-Stratonovich transformations. In section III 
we lay out our basic proposal and in section IV we provide a detailed Monte Carlo algorithm implementing this proposal. Subsequently,
in section V we present our results for the repulsive Hubbard model away from half filling and conclude in section VI.

\section{Functional integral formulation for many body theories}
The Hamiltonian for a general many body system with up to two body interactions is given by, 
\ba
\label{ham}
\mathcal{H} &=& \sum_{ij} \epsilon_{ij} a^{\dag}_i a_j + \sum_{ijkl} V_{ijkl} a^{\dag}_i a^{\dag}_j a_l a_k \\
  &=& \mathcal{K} + \mathcal{V},
\ea
where $a^{\dag}_i$ ($a_i$) creates (annihilates) a fermion in state $i$.
The thermal expectation value of an observable $\mathcal{X}$ at temperature $T$ ($ = 1/\beta)$ is calculated as
\be
\langle \mathcal{X} \rangle = \frac{\mbox{Tr~} [\mathcal X e^{- \beta \mathcal{H}}]}{\mbox{Tr~} [e^{- \beta \mathcal{H}}]}.
\ee  
With the help of the Suzuki-Trotter decomposition, the density operator can be written as
\be
\label{st}
e^{-\beta \mathcal{H}} = \left ( e^{-\Delta \beta \mathcal{K}}  e^{-\Delta \beta \mathcal{V}} \right )^{\beta/\Delta \beta} 
\ee
up to a discretization error that vanishes as $\Delta \beta \to 0$.

An arbitrary two body interaction operator $\mathcal{V}$ can be written as 
\be
\mathcal{V} = -\sum_{\alpha} \mathcal{O}_{\alpha}^2  + \mbox{1-body terms}\;,
\ee
where the operators $\mathcal{O}_{\alpha}$ are linear combinations of the bilinears in the creation and annihilation operators,
\be
\label{bilin}
\mathcal{O}_{\alpha} = \sum_{ij} (\lambda^{\alpha}_{ij} a^{\dag}_i a_j +  \eta^{\alpha \star}_{ij} a^{\dag}_i a^{\dag}_j + \eta^{\alpha}_{ij} a_i a_j )
\ee 
The partition function can now be written as a functional integral by introducing auxiliary fields for each operator $\mathcal{O}_{\alpha}$ 
and each imaginary time slice in Eq.~(\ref{st}) with the help of a continuous Hubbard-Stratonovich transformation (HST) 
\footnote{In principle, discrete HSTs can also be constructed, and are used quite extensively. But, we will not be concerned 
with them in this work.}
\be
\label{hs}
e^{\mathcal{O}^2/2} \propto \int d\phi \; e^{-\frac{\phi^2}{2} + \phi \mathcal{O}} 
\ee
where $\mathcal{O}$ is a generic operator. By doing so the interacting many body problem is 
converted to a problem of \emph{non-interacting} fermions with background auxiliary fields.
The price to be paid is that one now needs to integrate over all allowed values of the auxilary fields $\phi$.

The fermionic degrees of freedom can be integrated out (traced over)
exactly resulting in the following expression for the expectation value of an observable
\be
\label{obser}
\langle \mathcal{X} \rangle = \frac{\int_{\mathbb{R}^n} \mathcal{D} \bm{\phi} \; \mathcal{X} [\bm{\phi}] e^{-\mathcal{S}[\bm{\phi}]} }{\int_{\mathbb{R}^n} \mathcal{D} \bm{\phi}\;  e^{-\mathcal{S}[\bm{\phi}]}}
\ee
where the action $\mathcal{S}$ is
\be
\label{act}
\mathcal{S}[\bm{\phi}] = \sum_{\alpha \nu} \frac{\phi_{\alpha \nu}^2}{2} - \log \det \bm{M} [\bm{\phi}] \; .
\ee
and $\nu$ is the time slice index. 
The determinant of the matrix $\bm{M}$ is obtained after integrating out the fermionic degrees of freedom. 

The bilinear forms $\mathcal{O}_{\alpha}$ are not unique; there
is considerable freedom in choosing them. 
However, barring some special cases, an HST decoupling
\emph{cannot be found} such that $\mathcal{S}[\bm{\phi}]$ is  
real, i.e., the weight $e^{-\mathcal{S}[\bm{\phi}]}$ is positive semi-definite, for all $\bm{\phi}$.
 This is the source of the \emph{sign problem}. 

In the presence of the sign problem, $e^{-\mathcal{S}[\bm{\phi}]}$ cannot be treated
as a probability density, and hence Monte Carlo sampling cannot be used to evaluate ratio of the integrals in Eq.~(\ref{obser}).
In principle, it is possible to interpret $e^{- \Re \mathcal{S}[\bm{\phi}]}$ as the probability density and 
include $e^{ -i\Im \mathcal{S}[\bm{\phi}]}$ into the calculation of the observables. This reweighting method
is sometimes successful when the sign problem is not particularly severe (high temperature, small systems). But 
with increasing system size or decreasing temperature, the right hand side in Eq.~(\ref{obser}) is reduced to a ratio
exponentially small quantities with large variances, thus rendering these simulations unfeasible. 

\section{Quantum corrections via integration on the paths of steepest descent}

The smallness of the average sign $\langle e^{-i \Im S} \rangle$ is a result of delicate cancellations between the
contributions from the different regions of the phase space. On the other hand, analytic and semi-analytic calculations 
based on the functional integral formulation are usually based on the assumption that
in order to recover the relevant physics it is sufficient to consider the contribution from the configurations 
near a single dominant saddle point of the action (mean field solution). 

In these methods one (at least formally) goes through the following steps.
(i) The dominant mean field solution (saddle point) of the action are found; this saddle point may or may not lie in the 
original domain of integration ($\mathbb{R}^n$). (ii) The domain of integration is shifted to pass through the saddle point
and lie along the directions of statonary phase/steepest descent of the action. (iii) At the lowest order the integrals are simply replaced
by their values at the saddle point. (iv) Small quantum fluctuations are then included by expanding the action up to quadratic 
order at the saddle point and integrating along the directions of steepest descent for this approximate quadratic action. 

It is natural to conjecture, then, that \emph{full quantum corrections} about a mean field solution can be accounted for 
by replacing step (iv) above by an integration along the directions of steepest descent of the \emph{full action}. Next,
we provide further motivation for this conjecture, its implication for the sign problem and provide a concrete procedure,
based on stochastic sampling, for calculating the quantum corrections, so defined.

Assume, for the time being, that the action $\mathcal{S}[\bm{\phi}]$ is analytic for all $\bm{\phi} \in \mathbb{C}^n$, that the 
integrals in Eq.~(\ref{obser}) are convergent, and that all its  saddle points are non-degenerate.
In that case, one of the main results of Morse theory states that the integrals in Eq.~(\ref{obser}) can be reproduced exactly 
by replacing the integration over the real domain $\mathbb{R}^n$ by integrations over curved complex $n$-dimensional manifolds,
$\mathcal{J}^{n}_{\sigma}$, called the Lefschetz thimbles \cite{Pham1983,*Witten2010a,*Witten2010b},
\be
\label{rep}
\int_{\mathbb{R}^n} = \sum_{\sigma} \bm{n}_{\sigma} \int_{\mathcal{J}_{\sigma}^n}.
\ee
This result is basically the generalization of the contour shift method for one variable integrals.
The Lefschetz thimbles are the many variable analogues of the paths of steepest descent. In fact,
each Lefschetz thimble $\mathcal{J}^n_{\sigma}$ is attached to a saddle point $\bm{\phi}^0_{\sigma}$
\be
\left . \frac{\partial \mathcal{S}}{ \partial {\bm{\phi}}}\right \vert_{\bm{\phi} = \bm{\phi}^0_{\sigma}} = 0
\ee
and is the union of all paths of steepest descent which asymptotically end at the saddle point at $\tau \to \infty$; 
the paths of steepest descent are given by the solutions of the equations 
\be
\label{sd}
\frac{d\bm{\phi}}{d \tau} = -\overline{\frac{\partial \mathcal{S}}{\partial{\bm{\phi}}}} \;
\ee
where the overline represents complex conjugation.
(Please note that the parameter $\tau$ is in no way connected with the imaginary time.)
The integers $\bm{n}_{\sigma}$ are the intersection numbers between the hypersurface generated by the
paths of steepest ascent and the original domain of integration, in this case $\mathbb{R}^n$.

The Lefschetz thimbles are extremely attractive from the perspective of stochastic sampling 
because of two important properties (both of which can be easily verified from Eq.~(\ref{sd}))
\begin{enumerate}
	\item[(i)] the imaginary part of the action, $\Im \mathcal{S}[\bm{\phi}]$,
is a constant on a thimble, i.e., \emph{there is no sign problem due to the action on a thimble}
and 
\item [(ii)] the weight, $e^{- \Re \mathcal{S}[\bm{\phi}]}$, on a thimble is maximally localized near the saddle 
point, i.e. \emph{stochastic sampling is maximally effective on the thimble}.
\end{enumerate}

The presence of spontaneous breaking of continuous symmetries can lead to degenerate saddle points
($\det \partial^2_{\bm{\phi}} \mathcal{S} =0$). This degeneracy can be lifted by introducing an explicit
symmetry breaking term, $\epsilon \chi^{\dag}(\bm{\phi}) \chi(\bm{\phi})$, where $\chi(\bm{\phi})$ is 
the eigenvector of $\partial^2_{\bm{\phi}} \mathcal{S}$ with a zero eigenvalue.
 
Multiple calculations should be performed with successively 
smaller values of $\epsilon$, and the limit $\epsilon \to 0$ should be taken numerically.
One of the directions of steepest descent at finite $\epsilon$ becomes, in the limit
of vanishing $\epsilon$, the Goldstone mode of the broken symmetry. In what follows, we  
assume that procedure has been followed and the remaining saddle points are all non-degenerate.	

The presence of the logarithm of the fermionic determinant renders the action singular at a set of points
where $\det \bm{M}[\bm{\phi}]=0$. This is a set of measure zero. But, now the steepest descent paths can flow to 
 these points, instead of going all the way out to infinity. Note that, the branch cut in the 
logarithm does not introduce any additional complications because  $\Im \mathcal{S}$ is 
well defined on the thimble.

In general, the action can have a large number of saddle points in $\mathbb{C}^n$. 
It is not possible to find each of these saddle points and their intersection
numbers, much less to perform Monte Carlo integration on each one of them. However,
as we discuss below, this may not be necessary in many interesting cases. 

For theories with a well defined and unique continuum limit (e.g., lattice QCD, statistical systems in the vicinity of
a critical point, dilute atomic gases at unitarity), the universal properties of the model are expected to be a unique
outcome of (i) the symmetries, (ii) the degrees of freedom and (iii) the locality of the interactions. This is the statement 
of universality; it is not a theorem but is a necessary assumption that needs to be made in order to construct discretized 
versions of continuum theories.

In Refs.~\onlinecite{Cristoforetti2012, Cristoforetti2013}, 
it was argued that if the properties (i)-(iii) are preserved on a single thimble,
then the functional integral on that single thimble is a valid regularization of the original continuum theory and, on the basis of universality 
we should expect the universal properties obtained from this alternate regularization to be identical to those of the original formulation.
Thus for these theories it is clear which thimble should be chosen: the one where the properties (i)-(iii) are preserved.

This argument based on universality is clearly not sufficient for models which do not have a unique continuum limit; they essentially include
all models for many body systems away from criticality. In this paper we will be concerned with these models. 

Let us make the following definitions
\ba
\mathcal{Z}_{\sigma} &=& \int_{\mathcal{J}^n_{\sigma}} \mathcal{D} \bm{\phi} \; e^{-\Re \mathcal{S}[\bm{\phi}]} \\
\langle \mathcal{X} \rangle_{\sigma} &=& \frac{1}{\mathcal{Z}_{\sigma}}  \int_{\mathcal{J}^n_{\sigma}} \mathcal{D} \bm{\phi} \; \mathcal{X} [\bm{\phi}] e^{-\Re \mathcal{S}[\bm{\phi}]} 
\ea 
Then Eq.~(\ref{obser}) is replaced by,
\be
\label{obser1}
\langle \mathcal{X} \rangle = \frac{\displaystyle \sum_{p \in \mathcal{P}} \bm{n}_p \sum_{\sigma \in \mathcal{G}_p } e^{-i \Im \mathcal{S}[\bm{\phi}^0_{\sigma}]} \langle \mathcal{X} \rangle_{\sigma}  \mathcal{Z}_{\sigma}}{\displaystyle \sum_{p \in \mathcal{P}} \bm{n}_p \sum_{\sigma \in \mathcal{G}_p}  e^{-i \Im \mathcal{S}[\bm{\phi}^0_{\sigma}]} \mathcal{Z}_{\sigma}} 
\ee
where $\mathcal{P}$ is a set of phases, and $\mathcal{G}_p$ is the set of all thimbles belonging to the phase $p$. 
If $p$ is a symmetry unbroken phase, then it will have a unique mean field solution
and $\mathcal{G}_p$ will have a single member. In the case of a phase with a broken \emph{discrete} symmetry, 
$\mathcal{G}_p$ will contain all the thimbles connected by the generator of the symmetry which is broken in $p$.
In the above equation we have used the fact that the intersection numbers for thimbles belonging to the same phase are all equal.

Now, we make the following assumptions.
\begin{enumerate}
\item[(i)] The total number of `contributing' thimbles grow sub-exponentially with the number of degrees of freedom (dof). This can 
happen either if the total number of saddle points grow sub-exponentially with the number dofs, or if the total number of
saddle points grow exponentially with the number of dofs but the contributions from most of them cancel leaving
only a sub-exponential number of them contributing to the final integral.

Although this assumption is physically motivated, we cannot prove it from the consideration of the standard functional integral. In principle, it may be possible to 
construct actions where an exponentially large number of sub-dominant thimbles can overwhelm the contribution from the dominant thimble. We expect 
this not to be the case in the generic situations we are interested in.

\item[(ii)] We either know the dominant phase (the one with the largest $\mathcal{Z}_{\sigma}$) or we can narrow it down to a small number of candidates.     
      Note that, in order to identify the dominant phase it is not sufficient to compare the values of $\mathcal{S}[\bm{\phi^0_{\sigma}}]$; 
we also need to include quantum corrections to make this comparision, and the quantum corrections can be calculated only 
by performing the integration on the thimble.

In most interesting models of strongly correlated systems, the dominant saddle point is not known \emph{a priori}. 
However, in many cases it is possible to identify multiple candidates
by solving the mean field equations with certain imposed symmetries based on general considerations.

\item[(iii)] The thimble(s) belonging to the dominant phase (or potential candidates) are attached 
to `physical' mean field solutions,
i.e., that the corresponding mean field free energy is real. The mean field free energy is given by 
\be
\label{fmf}
\mathcal{F}^{\sigma}_{\rm MF} = \Re \mathcal{S}[\bm{\phi^0_{\sigma}}] - \log |\bm{n}_p| + i \{\Im \mathcal{S}[\bm{\phi^0_{\sigma}}] - \arg(\bm{n}_p)\}
\ee
The reality condition on $\mathcal{F}^{\sigma}_{\rm MF}$ implies that $\Im \mathcal{S}[\bm{\phi^0_{\sigma}}] - \arg(\bm{n}_{\sigma})=0$, 
i.e., $\Im \mathcal{S}[\bm{\phi^0_{\sigma}}]$ can only take discrete values $[0,\pi]$. 
\end{enumerate}

Each $\mathcal{Z}_{\sigma}$ in Eq.~(\ref{obser1}) is the exponential of an extensive quantity 
(the free energy on the thimble). Once we assume (i), it is easy to see that in the thermodynamic limit
only the $\mathcal{Z}_{\sigma}$ belonging to the dominant phase will contribute, the thimbles belonging 
to the other phases give vanishing contributions. The outer
sum (over $p$) is replaced by a single element, which implies that $\bm{n}_p$ in the numerator cancels with that in the denominator; $\langle \mathcal{X} \rangle$ 
does not depend on the $\bm{n}_p$ (by definition, $\bm{n}_p \neq 0$ in the dominant phase). It is sufficient to set $|\bm{n}_p|=1$ for all the phases.

We will restrict the outer sums over $p$ in Eq.~(\ref{obser1}) to the candidates for the  the dominant phase.
It follows from assumption (iii), this sum will not have any sign problem due to the action.

The Lefschetz thimble is a curved manifold in complex field variables.
Configurations on the thimble can be sampled by first defining a map between the 
field variables on the thimble and $\mathbb{R}^n$. This map can either be local \cite{Cristoforetti2012} or global \cite{Mukherjee2013}.
Stochastic sampling is then performed on the mapped real variables.

The Jacobian of the transformation between the  mapped real variables and the field 
variables,  $e^{\mathcal{R}[\bm{\phi}]}$, should be included as a 
reweighting factor in the calculation of the observables. One might suspect that the \emph{residual phase}, $\Im \mathcal{R}$,
\footnote{When a local map is used, $\mathcal{R}$ is a pure phase, $\Re R =0$ \cite{Cristoforetti2014}.}
might cause a sign problem.
However, there are very good reasons to believe that this is, in fact, not the case. 

In asymptotic expansions of the thimble around the saddle point $\langle e^{i \Im \mathcal{R}} \rangle =1$. 
The residual phase, $\Im \mathcal{R}$, can deviate substantially 
from zero only for configurations far away from the saddle point. Given, the nature of the thimble these configurations are maximally suppressed. 

The orientation of the thimble smoothly interpolates between the directions of steepest descent at the saddle point (determined 
by the quadratic part of the action) and the asymptotic directions of convergence; the residual phase can never oscillate unpredictibly,
it can only change smoothly. In short, the thimble achieves a tight correlation between the weight and the phase, which is precisely 
what is missing from the usual formulation leading to the sign problem.

Quantitative support for the above qualitative arguments is provided by the explicit calculation of the residual phase for a non-trivial
model with a severe sign problem \cite{Fujii2013}. There $\langle e^{i \Im \mathcal{R}} \rangle$ was found to be systematically larger 
than $0.99$ for all parameter values studied.

In a phase with broken symmetry, it is sufficient to sample field configurations on a single thimble. Configurations
on all other thimbles in the same phase can be generated by applying the generator of the said symmetry. Putting together
the considerations above we are led to the following expression for the $\langle \mathcal{X} \rangle$ in our
formulation
\be
\label{tobser}
\langle \mathcal{X} \rangle = \frac{\displaystyle \sum_{p \in \mathcal{P}} \mathcal{N}_p \int_{\mathcal{J}^n_p} \mathcal{D} \bm{\eta} \;  e^{\mathcal{R}[\bm{\phi}]}  \langle \mathcal{X} [\bm{\phi}] \rangle_p e^{-\Re \mathcal{S}[\bm{\phi}]}}{\displaystyle \sum_{p \in \mathcal{P}} \mathcal{N}_p \int_{\mathcal{J}^n_p} \mathcal{D} \bm{\eta} \;   e^{\mathcal{R}[\bm{\phi}]} e^{-\Re \mathcal{S}[\bm{\phi}]}}
\ee
where, $\mathcal{J}^n_p$ is the thimble attached to \emph{one} arbitrary saddle point $\bm{\phi}^0_p$ in the phase $p$, 
$\mathcal{D} \bm{\eta}$ is the real measure on the thimble,
$\mathcal{N}_p$ is the number of saddle points in the phase $p$ and $\langle \mathcal{X} [\bm{\phi}] \rangle_p $ is the average of the observable $\mathcal{X}[\bm{\phi}]$ over all the field configurations that are connected to $\bm{\phi}$ by the generator of the discrete symmetry broken in phase $p$.

\section{The Monte Carlo algorithm}

		We  propose the following Lefschetz thimble Monte Carlo algorithm for many body theories.
		\begin{enumerate}
			\item  Given a model for a many body system defined by a Hamiltonian as in Eq.~(\ref{ham}), construct the corresponding
				functional integral formulation with the help of a \emph{continuous} HST. There is considerable freedom in choosing the 
				HST and our method should be applicable to any of them. However, it is reasonable to expect that the `best' HST is determined 
				by a compromise between calculational convenience and how much of the symmetries of the original fermionic Hamiltonian can be retained 
				in the final action which defines the functional integral.
			\item Identify the dominant phase(s), and the associated saddle point(s) for the given parameter regime.
				The dominant saddle point(s) need not lie in the original domain of integration. 
In case of  degenerate saddle points due to a broken continuous symmetry, lift the degeneracy with an explicit symmetry breaking term.
  
				In cases, where the dominant phase is well known, our formulation can be used to obtain very precise 
				results by including all quantum corrections. In other cases, this formulation provides a well defined 
				setup for studying competing phases.
			\item    At the saddle point(s) $\bm{\phi}^0_p$, calculate the Hessian of the action 
                                 $\partial^2_{\bm{\phi}} \mathcal{S} [\bm{\phi}^0_p]$ and 
				 determine the directions of steepest descent from the generalized eigenvectors of the Hessian 
                                 as discussed in Ref.~\onlinecite{Mukherjee2013}.
                                This step needs to be done only once at the beginning of the simulation.
				The steepest descent directions at the saddle point(s) provide boundary conditions for Eq.~(\ref{sd}). 
			\item   If a single phase is considered then it is sufficient to generate configurations on a single thimble. 
                                In order to do this any of the algorithms described in Refs.~\onlinecite{Cristoforetti2012, Mukherjee2013} can be used.
 
                                If multiple phases are considered, then we need to incorporate a mechanism to jump between thimbles belonging 
				to different phases. In order to do this we will use a combination of Gibbs sampling and the algorithm 
                                described in Ref.~\onlinecite{Mukherjee2013}.

                                In this case, the variables $\bm{\eta}$ in the measure 
                                are the coordinates on the `Gaussian' thimble corresponding to the 
quadratic approximation of the action at the saddle point. Equivalently, they denote a direction and a length for integrating the Eq.~(\ref{sd}).
The phase and the fields are uniquely determined by the set $(p,\bm{\eta})$.

                                Start the sampling from any phase $p$. New configurations can be  sampled by repeating the following two steps
                                \begin{itemize}
                                  \item Given the phase $p$ (which also fixes the thimble), sample $\bm{\phi}$ from $e^{-\mathcal{S}[\bm{\phi}]}$
                                        using the algorithm described in Ref.~\onlinecite{Mukherjee2013}. 
					This fixes $\bm{\eta}$. Calculate the fields $\bm{\phi}_p$ on all the thimbles given $\bm{\eta}$.
  			          \item Given the $\bm{\phi}_p$ sample $p$ from $\mathcal{N}_p e^{-\mathcal{S}[\bm{\phi}_p]}$. 
                                \end{itemize}
			\item For each decorrelated configuration calculate the residual phase $\mathcal{R}$ and the observables
                              $\langle \mathcal{X} [\bm{\phi}] \rangle_p$. In broken symmetry phases, the latter can be calculated by 
				generating configurations on all the thimbles belonging to the phase with the help of the symmetry operation.
			\item Finally, calculate the expectation values of the observables by reweighting with $e^{\mathcal{R}}$.
	  \end{enumerate}

\section{Results for the Hubbard model away from half filling}
\begin{figure}[htbp]
  \includegraphics[width=\columnwidth]{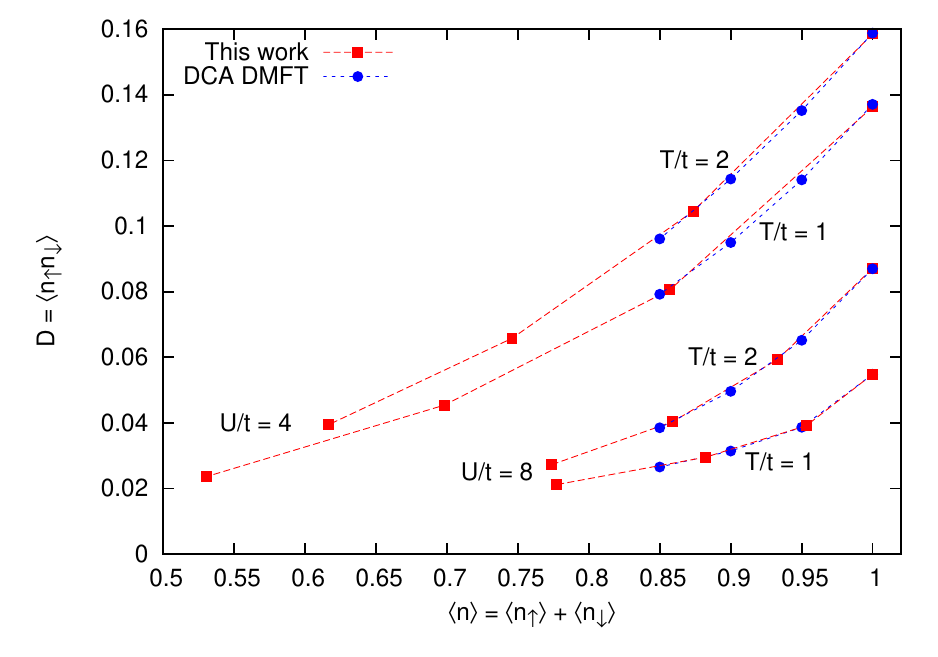}
  \caption{\label{dou} (Color online) Double occupation vs number density for the repulsive Hubbard model. Our results are shown with filled red squares, while the 
results from DCA DMFT \cite{Leblanc2013} are shown with filled blue circles. The error bars, where not visible, are smaller than the size of the symbols. 
The lines are meant as a guide to the eye.}
\end{figure}

\begin{figure}[htbp]
  \includegraphics[width=\columnwidth]{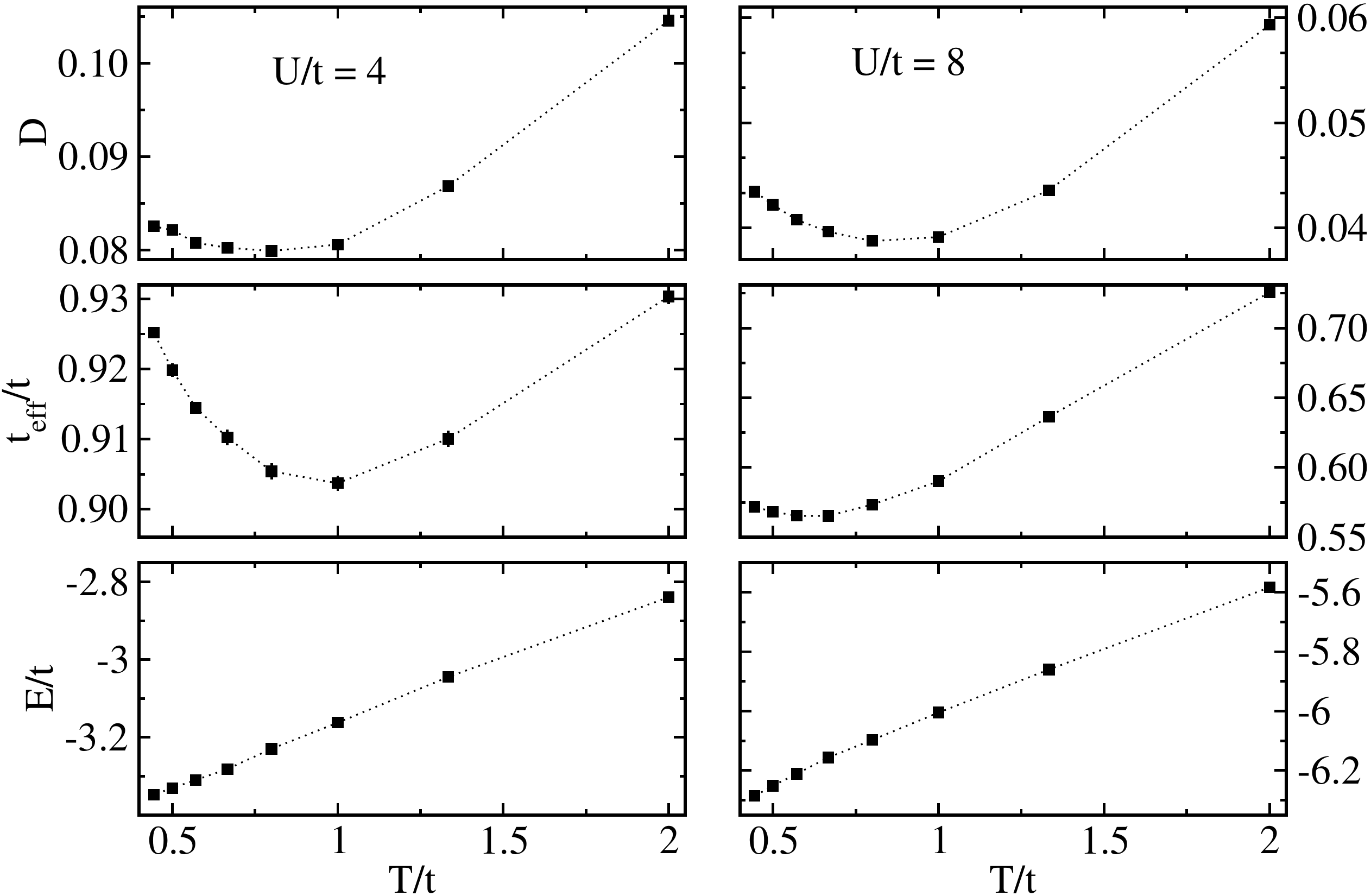}
  \caption{\label{mu10} Double occupation and effective hopping parameter vs temperature at $\mu/t=-1.0$ for the repulsive Hubbard model. 
The errorbars, where not visible, are smaller than the size of the symbols. The lines are meant as a guide to the eye. }
\end{figure}

In this section we will present our results for the Hubbard model. The Hubbard Hamiltonian is
\ba
\mathcal{H} &=& -t \sum_{\langle ij \rangle \sigma} (c^{\dag}_{i \sigma} c_{j \sigma} +c^{\dag}_{j \sigma} c_{i \sigma}  ) -  \sum_{i\sigma} \mu_{\sigma} n_{i\sigma} \nonumber \\
     &+& U \sum_{i} \left ( n_{i \uparrow } -\frac{1}{2} \right )  \left ( n_{i \downarrow} - \frac{1}{2} \right ) \\
		 &=& \mathcal{K} + \mathcal{V}
\ea
where $c^{\dag}_{i \sigma}$ ($c_{i \sigma}$) creates (destroys) a fermion of spin $\sigma$ at the lattice 
site $i$, $n_{i \sigma} = c^{\dag}_{i \sigma} c_{i \sigma}$ and $\langle ij \rangle$ denotes nearest neighbor sites. 
Also, $t$, $U$ and $\mu_{\sigma}$ are, respectively, the hopping parameter, the on site interaction strength and the chemical potential for the $\sigma$ spins.  

The interaction term can be written in the following manner
\ba
\label{iden}
n_{i \uparrow} n_{i \downarrow} &=& \frac{1}{2} \xi_i\left [ (e^{i \theta_i}n_{i \uparrow} + e^{-i\theta_i} n_{i \downarrow})^2 - (e^{2i\theta_i} n_{i \uparrow} + e^{-2i \theta_i} n_{i \downarrow})   \right ] \nonumber \\
                                &+& \frac{1}{2} (1-\xi_i)(e^{i \theta'_i} \Delta^{\dag}_i + e^{-\theta'_i} \Delta_i)^2	
\ea
where $\Delta_i = c_{i \uparrow} c_{i \downarrow}$. The interaction can now be decoupled by introducing auxiliary fields via HSTs for
each squared term in the above equation. The  $\theta_i$, $\theta'_i$ and $\xi_i$ can be arbitrary, illustrating the ambiguity in the HSTs - as noted earlier.
In fact, this is not even the most general decomposition of the interaction into a sum of squares of fermion bilinears and one body terms. 

No matter which decoupling HST is used, there is a sign problem for all parameter values except for certain special cases:
the repulsive case at half-filling ($ U > 0, \; \mu_{\sigma} = 0$)  and  the attractive case 
with spin balance ($U < 0, \;\mu_{\uparrow}=\mu{\downarrow}$) \cite{Batrouni1993}.
Here, we study the repulsive Hubbard model ($U > 0$) away from half filling ($\mu < 0$).

We use $\xi_i =1$ and $\theta_i=\pi/2$ in Eq.(\ref{iden}). 
Thus the auxiliary fields will couple to the on site magnetization $n_{i \uparrow} - n_{i \downarrow}$.
In this case, for $\bm{\phi} \in \mathbb{R}^n $,  $\det \bm{M} [\bm{\phi}] $ is real; but it can be positive or negative. 
This means that $\Im \mathcal{S}[\bm{\phi}]$ can only take discrete values $[0,\pi]$.

We present results for the intermediate coupling ($U/t=4$) and the strong coupling ($U/t=8$) regime. For each set of parameters we consider 
a single phase associated with uniform time independent real mean field solutions. The results presented here are limited to temperatures $T > 0.4t$.
Below these temperatures the dominant mean fields are most likely non-uniform. 

For a real saddle point $\partial^2_{\bm{\phi}} \mathcal{S}$ is a real matrix, and the directions of steepest descent are along $\mathbb{R}^n$.
Also, Eq.~(\ref{sd}) preserves the reality of the $\bm{\phi}$ fields. Therefore, the Lefschetz thimble is a subsector of the original domain
of integration: it is the region around the saddle point bounded by $\det \bm{M} [\bm{\phi}]=0$.

The action is invariant under the transformation, $\bm{\phi} \to - \bm{\phi}$. This symmetry is broken in the mean field theory, leading
to two saddle points related by the symmetry. We sample only on one of the thimbles and generate configurations on the other as discussed 
in the previous section.

Since the thimble is simply a subsector of $\mathbb{R}^n$, the sampling methods discussed in (4) of the previous section are not necessary. 
Instead, we use hybrid Monte Carlo with a leap-frog integrator for sampling \cite{Neal2011}. The step-size is kept small enough to prevent the trajectories
from crossing the zeros of $\det \bm{M} [\bm{\phi}]$. The residual phase is identically zero, i.e., there is no sign problem.

The non-ergodicity of hybrid Monte Carlo due to the inability to cross the zeros of $\det \bm{M} [\bm{\phi}]$ is usually considered to be an
undesireable feature (Note that this can happen even if there is no sign problem, i.e., when $\det \bm{M} [\bm{\phi}]$ is positive semi-definite
but not positive definite) \cite{White1988a,White1988b}. This conclusion would be absolutely correct if one wanted to explore the whole phase space. 
However, we \emph{want to} stay in a single (special) subsector of the phase space. And it is precisely this non-ergodicity of the method that 
we exploit to achieve that.

In Fig.~(\ref{dou}) we show the double occupancy as a function of the number density for different temperatures and interaction strengths. 
Double occupancy has been proposed as an experimentally  accessible probe for signatures of antiferromagnetism \cite{Gorelik2010,*Gorelik2012}. 
It has also been suggested that this observable can be used for thermometry in optical lattices in the high temperature regime ($T/t > 1$) \cite{Paiva2010}.  

The double occupancy is an extremely local observable and it reaches its thermodynamic limit very quickly as a function of the lattice size.
Our results shown in Fig.~(\ref{dou}) are for an $8^2$ lattice, but we have checked in some selected cases that the results for $10^2$ and 
$12^2$ lattices are within $\sim 1\%$. 

At half-filling, $\langle n\rangle = 1$, our results are in perfect agreement with the latest DQMC results \cite{Paiva2010,Scalettar2014}. 
In Fig.~(\ref{dou}) we compare our results for different fillings with the recent DCA DMFT calculations~\cite{Leblanc2013,Leblanc2014}, which is expected 
to be accurate at these temperatures. There is excellent agreement between results obtained from these two methods.  

In Fig.~\ref{mu10} we show double occupancy as a function of temperature for $\mu=-1$. We see a slight enhancement of the double occupancy below $T/t < 1$. 
It was proposed in Ref.~\onlinecite{Werner2005} that this effect, which is much more pronounced in dynamical mean field theory, can be exploited to achieve 
interaction induced cooling for cold fermions in an optical lattice.  

We also show in Fig.~\ref{mu10} the effective hopping parameter defined as the ratio of the kinetic energy for the interacting system and that for the 
non interacting one, $t_{\rm eff}/t = \langle \mathcal{K} \rangle_U/\langle \mathcal{K} \rangle_0$. The enhancement of the double occupancy results in an increase 
of the potential energy at low temperature. This is offset by a decrease in the kinetic energy (note that the the kinetic energy in negative), resulting in a smooth decrease
of the total energy with decreasing temperature, as shown in the bottom panels in Fig.~\ref{mu10}.

 \section{Conclusion}
 A general solution to the sign problem in the conventional sense would require an \emph{exact} mapping of high dimensional oscillatory integrals
 to a problem which can be solved in polynomial time on a classical computer. This appears to be an NP (nondeterministic polynomial time) hard problem 
 \cite{Troyer2005}. However, this does not preclude the possibility of a reformulation which may not reproduce exact functional integral, but neverthess
 preserves the correct physics, at least for a large subset of physical systems. 

 In previous papers we have shown how ideas used in constructing asymptotic expansions can be generalized to construct an alternate regularization 
 of the path integral, which can then be used performing efficient Monte Carlo simulations for systems with a sign problem. The notion of universality 
 of the continuum limit was used as a primary mode of justification for this reformulation of quantum field theories.

 Here, we take these ideas a step further. We showed that even for many body systems with no unique continuum limit, the same ideas can be used to 
 calculate full quantum corrections. We argued that this represents a natural non-perturbative 
generalization of the ideas used in constructing mean field theories and its various extensions.

 We provided a detailed algorithm that can be used to implement these ideas numerically. As a preliminary application we studied the Hubbard model
 away from half filling at moderate temperatures. Our results are in excellent agreement with the latest state of the art calculations from
 cluster dynamical mean field theory.

 We believe that our method is the ideal platform for combining the physical insight gained from mean field theory and the power of quantum Monte 
 Carlo. It will be especially interesting and challenging to apply it to strongly correlated systems at low temperatures where the dominant phase of 
 the system is not known.

\section*{Acknowledgments}
We are grateful to L.~Scorzato for numerous valuable discussions and careful comments on the manuscript. We would like to thank R.~Scalettar and 
J.~LeBlanc for sharing the latest results from DQMC and DCA DMFT, respectively.
\bibliographystyle{apsrev4-1}
 \bibliography{hubb_thim}
\end{document}